\documentclass[usenatbib]{mn2e}

\usepackage{graphicx,subfigure,tikz}
\usepackage{lscape}
\usepackage{txfonts}
\usepackage{url}
\usepackage{lastpage}
\usepackage{enumerate}

\title[Abundance of satellites and morphology of host]
{The abundance of satellites depends strongly on the morphology of the host galaxy}

\author[Pablo Ruiz, Ignacio Trujillo \& Esther M\'armol-Queralt\'o]{Pablo Ruiz$^{}$ \thanks{E-mail:
ruihern@gmail.com}, Ignacio Trujillo$^{1,2}$ and Esther M\'armol-Queralt\'o$^{1,2,3}$\\
$^{1}$Departamento de Astrof\'{\i}sica, Universidad de La Laguna, E-38205 La 
Laguna, Tenerife, Spain\\
$^{2}$ Instituto de Astrof\'{\i}sica de Canarias, c/ V\'{\i}a L\'actea s/n, 
E-38205 La Laguna, Tenerife, Spain\\
$^{3}$ Institute for Astronomy, University of Edinburgh, Royal Observatory, Edinburgh EH9 3HJ, UK}

\begin{document}

\date{Accepted 2015 September 1. Received 2015 August 21; in original form 2015 March 30}

\pagerange{\pageref{firstpage}--\pageref{lastpage}} \pubyear{2015}

\maketitle

\label{firstpage}


\begin{abstract}

Using the spectroscopic catalogue of the Sloan Digital Sky Survey Data Release 10, 
we have explored the abundance of satellites around a sample of 254 
massive (10$^{11}<$ M$_\star<$2$\times$10$^{11}$M$_{\sun}$) local (z $<$ 0.025) galaxies.
We have divided our sample into four morphological groups (E, S0, Sa, Sb/c). We
find that the number of satellites with M$_\star\gtrsim$10$^{9}$M$_{\sun}$ and 
R$<$300 kpc depends drastically on the morphology of the central galaxy. 
The average number of satellites per galaxy host ($N_{\rm Sat}/N_{\rm Host}$)
down to a mass ratio of 1:100 is: $4.5\pm0.3$ for E hosts, $2.6\pm0.2$
for S0, $1.5\pm0.1$ for Sa and $1.2\pm0.2$ for Sb/c. The amount of stellar mass 
enclosed by the satellites around massive E-type galaxies is a factor of 2, 4 and
5 larger than the mass in the satellites of S0, Sa and Sb/c-types, respectively. 
If these satellites would eventually infall into the host galaxies, for all the 
morphological types, the merger channel will be largely dominated by satellites
with a mass ratio satellite-host $\mu>$0.1. The fact that massive elliptical
galaxies have a significant larger number of satellites than massive spirals 
could point out that elliptical galaxies inhabit heavier dark matter 
haloes than equally massive galaxies with later morphological types. If this hypothesis is
correct, the dark matter haloes of late-type spiral galaxies are 
a factor of $\sim$2-3 more efficient on producing galaxies with the same stellar mass
than those dark matter haloes of early-type galaxies. 

\end{abstract}

\section{Introduction}

Galaxy mergers have been raised in the last years as the most likely channel of
size and mass growth of massive (M$_\star\gtrsim$10$^{11}$M$_{\sun}$) galaxies
through cosmic time. Numerous observational and theoretical studies support this mode
of growth, a mechanism that has increased the size and mass of the massive galaxies
during the last $\sim$11 Gyr. In this scenario, the ancestors of the present-day
most massive galaxies created the bulk of their mass in a short but very intense 
starburst event at z$\gtrsim$2 \citep[e.g.][]{Keres2005,Dekel2009,Oser2010,Ricciardelli2010,
Wuyts2010,Bournaud2011} having, in that first evolutionary stage, a structure more compact. 
Later, a progressive process of mergers with satellites produced the envelopes
that we see today surrounding these galaxies \citep{Khochfar2006, Oser2010,Feldmann2011}. 
Many works support the above scheme, finding evidences for a continuous size evolution
of the massive galaxies since z$\sim$3 \citep[e.g.][]{Trujillo2007, Buitrago2008} 
mainly produced by the formation of the outer most regions ~\citep[e.g.]
[]{Bezanson2009, Hopkins2009a, Carrasco2010, vanDokkum2010, Montes2014}.
In addition, other authors have found that the average velocity dispersion of the massive galaxies 
have decreased mildly since z$\sim$2 as expected theoretically from the galaxy merger
scenario \citep[e.g][]{Cenarro2009}. Other observations point out that this size evolution 
of the massive galaxies does not depend on the age of the stellar population \citep{Trujillo2011}
nor on their intrinsic sizes \citep{Diaz2013}. This suggests a growth engine external to the
galaxy properties. The absence of a significant number of relic galaxies in the nearby Universe
also favours a merging scenario \citep[e.g.][]{{Trujillo2009, Trujillo2014}, Taylor2010}.

The above merging scenario can be alternatively probed measuring the satellite abundances around 
massive galaxies as cosmic time flows ~\citep[e.g.][]{Newman2012}. There are many works that 
have studied in detail the properties of the satellite galaxies over time~\citep{Jackson2010,{Nierenberg2011,Nierenberg2013},
Man2012, {Marmol-Queralto2012, Marmol-Queralto2013},Newman2012, Huertas2013, Ferreras2014}. 
In particular, ~\cite{Marmol-Queralto2012} found that the fraction of massive galaxies 
with satellites of a given mass ratio (1:100 up to $z=1$ and 1:10 up to $z=2$) have remained constant
with time. A behaviour which is in qualitative good agreement with semi-analytical 
predictions based on the $\Lambda$ cold dark matter ($\Lambda$CDM) model \citep{Quilis2012}. 
However, the semi-analytical models over-predict the fraction of massive galaxies with 
satellites down to 1:100 mass ratio by a factor of $\sim$2. 

Parameters such as the abundance of satellites, their  distribution or their intrinsic
properties are intimately bound up with their host merger histories. These properties
are thus, closely related to the  underlying cosmology and they can be used to establish
useful constrains to the models. The colours and 
structural properties of the host galaxies can be modified by gravitational interactions
with their satellites. The main goal of this work is to analyse the 
relation between the abundance of satellites and the host morphology in a sample
of nearby massive ($\sim$10$^{11}$M$_{\sun}$) galaxies. 
We segregate our sample of galaxies in four groups which are 
representative of different structural configurations. 
Our morphological classification identifies visually the groups E, S0, Sa and
Sb/c to explore the correlation between the number of satellites and
the morphology of the host. Our approach differs from previous studies based on
colours \citep{Chen2008,QuanGuo2011,Wang2012}, or those based on more general grouping (e.g. early-, late-
or spheroid- disc like) of the massive galaxies  \citep[e.g.][]{QuanGuo2011,Marmol-Queralto2012,Nierenberg2012}.  

Our samples of massive galaxies could be used in future works to test
the $\Lambda$CDM predictions about the number of satellites surrounding
the most massive galaxies in the present-day Universe \citep[see e.g.][]
{Chen2008,Liu2011,Wang2012} according to their morphology. Our study also allows us to
explore which is the most likely merging channel of present-day massive galaxies,
i.e. which type of satellites contribute most to the mass increase of their 
host galaxies in case they eventually merge with its main object. Consequently, 
this local study, along with other works at higher $z$ \citep[see e.g.][]
{Ferreras2014}, allows us to explore whether the merging channel has
changed with time and its dependence with the host morphology.
 
This paper is structured as follows. In Section ~\ref{sec:data}, we describe our
sample of hosts and satellite galaxies, their completeness and their stellar mass estimates. 
Section  ~\ref{sec:statistic_estimation} explains the satellite selection criteria and 
the methods used to clean our sample from background and clustering contamination. Our results
concerning the satellite abundances of the distinct samples are
presented in Section ~\ref{results}. Section ~\ref{sec:discussion} discusses 
the main results of this paper and finally our work is summarized in Section
~\ref{sec:conclusions}. Hereafter, we assume a cosmology with $\Omega_{\rm m}= 0.3$,
$\Omega_{\rm \Lambda}= 0.7$ and H$_0 = 70$ km~s$^{-1}$~Mpc$^{-1}$. 
\section{The data}\label{sec:data} 

In this paper, we use the `specphoto' spectroscopic catalogue of Sloan Digital Sky Survey (SDSS) Data
Release 10 \citep[DR10;][]{Ahn2014} 
to explore the abundance of satellites around a sample of massive galaxies
in the nearby Universe ($z\leq0.025$). The spectroscopic completeness of this catalogue is 
90 per cent down to $r=17.7$ mag. The 
catalogue includes a total of 1507954 Baryonic Oscillation Spectroscopic Survey (BOSS) spectra comprising 927844 galaxy
spectra, 182009 quasar spectra and 159327 stellar spectra selected over 
6373.2 deg$^{2}$. We select those objects labelled 
as `GALAXY' within the data set `specphoto'. This subset only has galaxies 
where the `SpecObj' is a `sciencePrimary' object, and  the BEST PhotoObj is
a PRIMARY.  

We structure this section as follows. First, we show the procedure to estimate 
the stellar masses of the galaxies of the catalogue. Then, we describe the 
selection of our host galaxies, the catalogue of potential satellites and 
finally, we study the completeness of our satellite population.

\subsection{The stellar mass estimation}\label{sec:sme}

One of the goals of this work is to analyse the abundance of satellites as a 
function of the mass ratio satellite--host. To estimate the stellar masses, we
use the \citet{Bell2003}'s recipe, assuming
a \citet{Kroupa2001} initial mass function (IMF). We take the `modelMag' $g$- and $r$-band
 magnitudes from the `specphoto' SDSS DR10 catalogue once they 
have been corrected from Galactic extinction \citep{Schlegel1998}. Then, we estimate
the mass-to-light ($M/L$) ratio from the rest-frame $g-r$ colour, being the $M/L$
ratio estimated in the $r$ band as follows:

\begin{equation}\label{formula_mass_lum}
{\rm log}(M/L)_{r} = a_{r} + b_{r} (g-r) - 0.15{\rm ,} 
\end{equation}

where $a_{\rm r}=-0.306$ and $b_{\rm r}=1.097$ are the coefficients applied 
for determining the M/L ratio in the r band and 0.15 is 
subtracted to match the results to a Kroupa IMF.

Using these computed $(M/L)_{\rm r}$, we can directly estimate the stellar masses 
using the next relationship:

\begin{equation}\label{formula_mass}
{\rm log}(M/M_\odot) = {\rm log}(M/L)_{r} - 0.4(M_{r} - M_{\odot,r}){\rm ,}
\end{equation}

where $M_{r}$ is the absolute magnitude of the galaxy and $M_{\odot,r}$=4.68
the absolute magnitude of the Sun in the SDSS $r$ band. Given that our study is focused
on objects at very low redshift, we do not apply $K$-corrections to the above $g$ and $r$ 
magnitudes since it will not affect significantly our mass estimates. In fact,
the expected values for $K$-corrections at $z<0.025$ are typically below $\sim$1 per cent
relative to calibration errors found for $g$ and $r$ filters in the photometry of SDSS
DR10~\citep{Padmanabhan2008}.

To test how reliable our stellar mass estimates based on colours are, we have compared
our stellar masses with those from \citet{Nair2010} based on \citet{Kauffmann2003}.
This comparison can be done only for a subset of the galaxies in our catalogue. As 
the result of this comparison we find a bias of 0.2 dex~\citet[being the][smaller]{Nair2010}
with a typical uncertainty of 0.1 dex among the two stellar
mass estimators. On what follows, we take that uncertainty as representative of our error 
estimation of the stellar mass. The above bias is not surprising taking into account the
different methodologies and stellar population models used in both estimates of the
stellar mass. 

\subsection{The sample of host galaxies}\label{sec:hosts}

\begin{figure}
\includegraphics[width=1.0\columnwidth,clip=true]{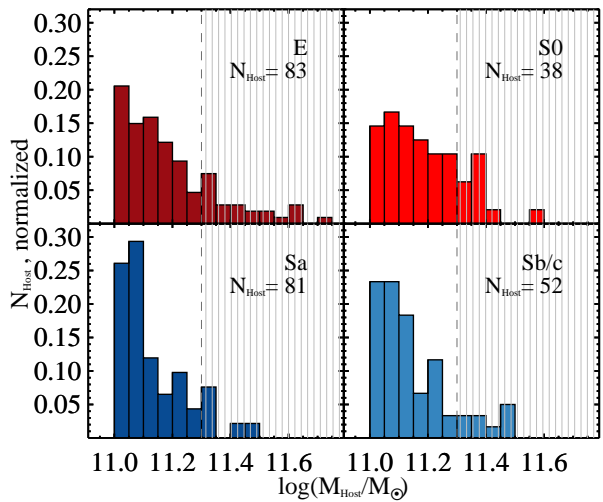}

\caption{
Distribution of the normalized stellar mass for the host of each morphological type. 
The upper left-hand panel corresponds to the massive E type, the upper right to S0 type, 
the lower left to Sa and the lower right to the Sb/c massive galaxies. The non-stripped 
area illustrates the distribution of massive 
galaxies chosen (10$^{11}<$M$_{\star}/$M$_{\sun}<$2$\times$10$^{11}$) for this work. The N$_{\rm Host}$ 
label in the panels indicates the number of host found within that stellar mass interval. 
} 

\label{host_distribution} 
\end{figure}

Using the available data, we estimate the stellar mass of our galaxies 
as we explained above (Section \ref{sec:sme}).
To build our sample of massive host galaxies, we select only the galaxies 
with M$_\star\gtrsim$10$^{11}$M$_{\sun}$.
We limit our sample to galaxies 
with $z<0.025$ (i.e. at a distance $\lesssim$100 Mpc). The average apparent
r band magnitude of our host galaxies is $r\sim$13 mag. This implies, 
taking into account the spectroscopic completeness limit of our catalogue,
that we can identify potential satellites with stellar masses 100 times 
less massive than their hosts. We have discarded massive hosts whose $M/L$
(as computed by their $g-r$ colour) were unreasonable (i.e. out of the 
interval $0.1<(M/L)_{r}<4$) and those ones with large photometric 
errors. The percentage of massive galaxies discarded by these reasons were 
5 per cent. We also check visually our sample of massive galaxies to reject 
objects wrongly labelled as galaxies as well as those ones in clear 
interaction with another galaxy. Also, to avoid incompleteness in the
number of satellites due to area coverage, 27 massive galaxies were 
additionally discarded due to their proximity to the edge of the SDSS DR10
spectroscopic footprint.

Finally, the mass distribution of the different morphological types was checked.
Our morphological segregation is explained in the next subsection.
The distribution of the mass of the hosts is illustrated in the Fig.~\ref{host_distribution}
and shows a high-mass-tail for the massive ellipticals which is not so prominent
for the rest of galaxy types. To avoid any potential bias caused by this
different mass distribution of the hosts, we established an upper mass limit of
2$\times$10$^{11}$M$_{\sun}$. Thus, our final host sample is composed of 254 massive galaxies. 

\begin{figure}
\includegraphics[width=1.0\columnwidth,clip=true]{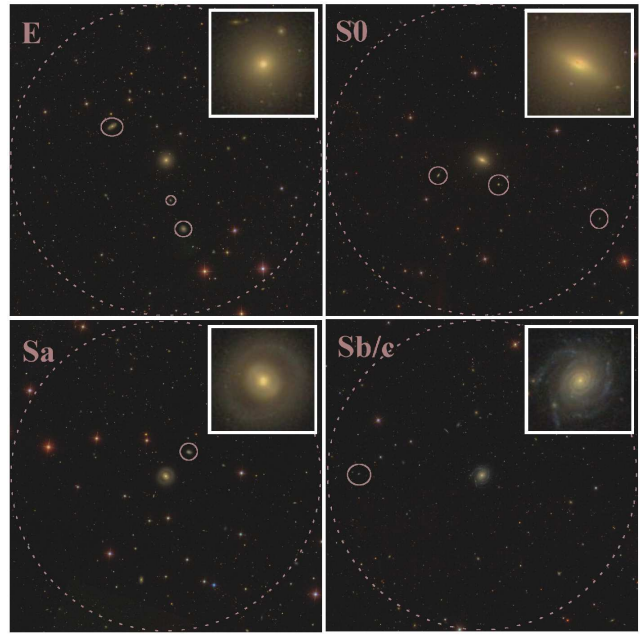}

\caption{
Four examples of massive galaxies in our sample and their surroundings. The 
dashed circles enclose a region of 300 kpc in radius. The small circles indicate
the position of the satellites. The upper right-hand panels are a zoom-in of the 
host galaxies to illustrate their morphologies. 
} 

\label{host_examples} 
\end{figure}

The main goal of this work is to study the local abundance of satellites as a 
function of the morphological type of the massive hosts. Unlike other studies whose 
separation is based on colours, we do our own visual classification based on the Hubble
classification using the SDSS DR10 images (see next subsection). Within the mass range
(10$^{11}<$M$_{\star}/$M$_{\sun}<$2$\times$10$^{11}$) we find 83 E-type,
38 S0-type, 81 Sa-type and 52 Sb/c-type galaxies. Fig.~\ref{host_examples} 
illustrates some of our host galaxies. Our visual classification is compared with 
the one done by \citet{Nair2010} for 89 massive galaxies we have in common. We 
find an agreement of 84 per cent.

\subsubsection{Sorting the galaxy hosts into morphological classes}

In this work we have segregated our galaxies into four different morphological types (E, S0,
Sa and Sb/c). The physical motivation for sorting into these four categories is related to the
expected strong connection between the evolutionary path followed by the galaxies and their
detailed morphologies. This connection leaves their imprints on the relation
between the specific angular momentum of the galaxies, at a fixed stellar mass, and 
the galaxy morphology \citep[see e.g.][]{Romanowsky2012}. This link can also be seen
in the different shape of the outermost regions of the galaxies depending on the global
morphological type \citep[e.g.][]{Pohlen2006,Erwin2008}. If as expected, the merging activity of the
galaxies is connected to the galaxy morphology, a natural prediction is that the number of
satellites surrounding the galaxies should be also related with the shape of their host
galaxies. It is worth noting that the merging activity is likely linked to the amount of
mass contained within the bulge of the galaxies \citep[e.g.][]{Hernquist1991, Steinmetz1995,
Fu2003,Zavala2008,Kroupa2010, Kroupa2012}. The prominence of the bulge is one of the key
ingredients in the galaxy morphological criteria, and consequently, a detailed segregation
among the morphological types (beyond a disc/elliptical separation) could be connected with
the number of satellites around the host galaxies.

To classify our galaxies we have followed the traditional Hubble classification scheme. For
all our galaxies, we look in detail the colour stamps provided for each of them by the finding
chart tool of SDSS. To disentangle among ellipticals and S0s, we search for any evidence
of a less steeply declining brightness in the outer region of the galaxies beyond the central
brightness condensation. In cases where the inclination of the disc component of the S0 galaxy
was clearly showing a flat outer structure or when some dust features were obvious, the
distinction between the two galaxy types was relatively simple. Among the disc galaxy
population, the segregation between Sa and Sb/c was done according to the relevance of the
bulge in producing the overall light distribution as well as exploring the properties of the
spiral arm structure, i.e. the tightness with which the spiral arms are wound and the number
of substructure visible in those features.

\subsection{The sample of satellite galaxies}\label{sub:sats}

To select our sample of potential satellites, we also use the `specphoto'
catalogue of SDSS DR10. Our satellites are those galaxies in the catalogue which
fulfil the proximity criteria (i.e. projected distance to the hosts)
and stellar mass criteria that we will explain in the next section (Section
~\ref{sec:statistic_estimation}). 

On building our sample of satellites, we have to prevent the inclusion
of objects with deficient measurement of its colour because this would lead
to a wrong estimate of its stellar mass. To conduct this task, we need 
to account for the photometric errors both in $g$ and $r$ bands to assure the 
colour is measured with enough confidence. For this reason, in addition to
the magnitude limit in the $r$ band we have used above, we also demand that 
the photometric error at estimating the number counts of each galaxy will
be less than 5$\sigma$ the expected error at measuring their number counts. 
In other words, acceptable photometric error for each object for us are 
those whose error(counts) is $\lesssim$5$\times\sqrt{(counts + \sigma_{\rm sky}^2)}$,
with $\sigma_{\rm sky}$ the uncertainty (in counts) at measuring the sky value in
each band. We find as typical values for the sky in the SDSS images 24.88 
counts ($g$ band) and 23.96 counts ($r$ band). We have used the following 
set of equations to transform our magnitudes and error(mag) provided by
the catalogues into counts and error(counts):

\begin{eqnarray}
{\rm mag=}-2.5\log\left(\frac{\rm counts}{\rm exptime}10^{0.4(aa+kk\times \rm airmass)}\right)\\ 
{\rm error(mag)=}\frac{2.5}{\ln
10}\frac{\rm error(counts)}{\rm counts}\end{eqnarray} 

with exptime=53.907 s and
$aa$, $kk$ and airmass provided for each object, being $aa$ and $kk$ the values of the
zero-point and the extinction coefficient respectively. Those galaxies in our 
catalogue which show a photometric error larger than those values (in any 
of the two bands) are discarded from the analysis since these ones could be 
linked to artefacts in the image, proximity to bright nearby companions, etc. We have
estimated the number of galaxies rejected because of large photometric errors,
finding that less than 0.5 per cent of the objects are discarded, a reasonable
result since these objects are relatively bright.

\subsubsection{The completeness of the satellite sample}
\label{sec:compspeccat}

\begin{figure}
\includegraphics[width=1.0\columnwidth,clip=true]{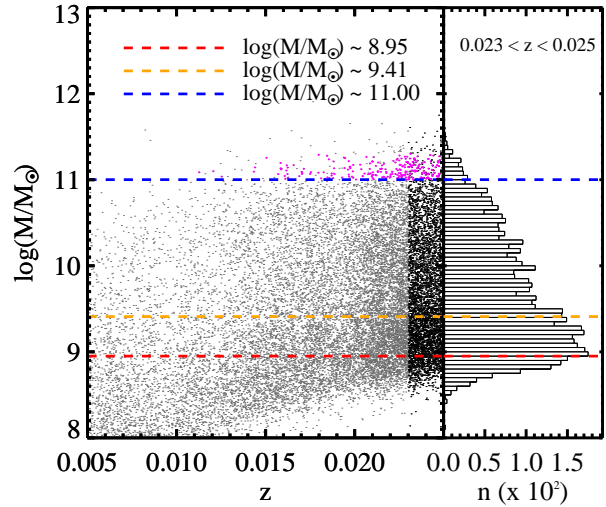}

\caption{Left-hand panel: the points in grey correspond to stellar mass distribution versus redshift
for all the galaxies in the `specphoto' SDSS DR10 catalogue. The pink points represent our sample of
massive host galaxies. The black points indicate those objects in the 
right-hand panel within the redshift (0.023$<$z$<$0.025) used to estimate the
stellar mass completeness. Right-hand panel: stellar mass distribution for the 
spectroscopic catalogue in the redshift interval 0.023$<$z$<$0.025.
The red dashed line is the completeness limit $\sim$8.9$\times$10$^{8}$M$\odot$ 
whereas the blue line represents  the minimum stellar mass for the sample of massive
galaxies (10$^{11}$M$\odot$). A conservative estimation for the
stellar mass completeness of the spectroscopic sample is showed by the
orange dashed line: $\sim$2.6$\times$10$^{9}$M$\odot$ (see text for details).} 

\label{spec_comp} 
\end{figure}

Once the stellar masses of the galaxies of our catalogue are determined,
we can estimate down to which stellar mass the catalogue is complete. 
The stellar mass limit for completeness is a function of the redshift (see 
Fig.~\ref{spec_comp}). To explore the degree of completeness 
of  satellites down to M$_\star\sim$10$^{9}$M$_{\sun}$, we have used our most 
distant redshift interval 0.023$<$z$<$0.025. The peak on the mass distribution
of the galaxies in the catalogue at z$\sim$0.024 is $\log(M_\star/M\odot$)$\sim$
8.95. If we now take into account that the minimum stellar mass that we have 
fixed for our hosts ($\log(M_\star/M\odot$)$\sim$11.0), we should be able to study
with completeness satellites whose M$_{\rm Sat}$/M$_{\rm Host}$ $\gtrsim$ 0.01.

However, it is worth noting that there is a potential bias to miss the oldest
satellites at a fixed stellar mass. This is because the catalogue is complete
in redshift down to a given apparent $r$-band magnitude ($r\sim$17.7). That value
translates into the following absolute magnitude for the sample at z=0.025: 
M$_{r}$=-17.65 mag. To transform this absolute magnitude into a stellar
mass limit we need to have an estimation of the stellar $M/L$ ratio of our 
satellites. To do this, we assume a conservative age for the less massive
galaxies of 10 Gyr. Using the MIUSCAT spectral energy distributions SEDs developed by \citet{Vazdekis2012}
and \citet{Ricciardelli2012}, a solar metallicity and a Kroupa IMF, the $(M/L)_{\rm r}$
ratio for these objects is around 3. This translates into the following stellar
mass: $\log(M_\star/M\odot$)$\sim$9.41 (i.e. a mass ratio satellite--host of 1:40).
On what follows, we will consider this value as the most conservative mass 
completeness limit of our satellite galaxies. 

\section{Satellite selection criteria}\label{sec:statistic_estimation}

Our criteria to search for potential satellites around the host sample are based on 
the next three steps.

(i) We detect all the galaxies in the 'specphoto' SDSS DR10 catalogue which
are within a projected radial distance to our central galaxies of $R=$300 kpc. 
We only consider those host galaxies when the area enclosed by the search radius
of satellites is fully contained within the catalogue borders. As we mentioned above,
27 hosts were discarded due to their proximity to the catalogue edge. 
Our adopted search radius of 300 kpc is a compromise between having a large 
area for finding a significant number of satellite candidates gravitationally 
bound to our central massive galaxies but not as large as to be severely 
contaminated by background and foreground objects (see Section ~\ref{sub:background}).

(ii) The absolute difference between satellite redshifts and the redshift of
the central galaxies must be lower than 1000 km s$^{-1}$. This value has been
used before in the literature to select gravitationally bound satellites of 
massive galaxies \citep[see e.g.][]{Wang2012,Ruiz2014}. The velocity distribution
of the galaxies around the massive host galaxies selected that way is close 
to a Gaussian shape with a dispersion of 300 km s$^{-1}$. Consequently, our 
criteria enclose the vast majority of satellites around the massive galaxies. 

(iii) The mass ratio
between our host massive galaxy and the satellite should be above 1:100.

Those objects which fulfil the above criteria are counted as potential satellites of their 
hosts. The number of satellites observed around a host sample is defined as N$_{\rm Obs}$.
Before showing our results, we will address the potential biases that can
affect our counting of satellites around the massive hosts. 

\subsection{Background correction}\label{sub:background}

Despite we have used spectroscopic redshift information to select our potential satellite
galaxies, there is still a fraction of objects that satisfy all the above criteria but are not
gravitationally bound to our massive galaxies. These objects are counted as satellites because the
uncertainties on their redshift estimates include them within our searching redshift range. These
foreground and background objects (hereafter we will use the term background to refer to both of them)
constitute an important source of uncertainty in this kind of studies. Consequently,
it is key to estimate accurately the background contamination in order to statistically subtract
its contribution from the number of galaxies hosting satellites.

To estimate the typical number of background objects that contaminates our satellite samples, 
we have developed a set of simulations. The procedure consists on placing a number of mock 
galaxies (equal to the number of our host galaxies) randomly through the volume of the catalogue
conserving their original values in the stellar mass and redshift of the sample of 
massive galaxies. Once we have placed our mock galaxies through the catalogue,
we count which number of these galaxies have fake satellites around them
taking into account the criteria of stellar mass, redshift and distance explained in the
above section. This procedure is repeated 2000 times to have a robust estimate of the number
of 'satellites' around the mock host galaxies. We define this average number as N$_{\rm S}$.
Then, being N$_{\rm Obs}$ the number of observed satellite galaxies around either of
our host massive galaxies, we correct statistically its excess subtracting the number of satellites 
representative of the background (N$_{\rm S}$), such as it is shown in the equation below
(equation ~\ref{formula_sim}). By construction, N$_{\rm S}$ is independent on the 
morphology of the host. 

\begin{equation}\label{formula_sim}
N_{\rm Sat,S} = N_{\rm Obs} - N_{\rm S}
\end{equation}

To take into account that the environmental density and the morphology are linked for 
massive nearby galaxies, we also compute a clustering correction as explained in the following 
section.

\subsection{Clustering correction}\label{sub:clustering}

At low redshift, the over density regions are specially populated by massive galaxies.
It is worth therefore exploring whether our background correction is representative
of the contamination of sources surrounding our host galaxies or whether it is 
necessary to compute the excess of probability of finding 'satellites' in these
environments. We term this as clustering. 

Being the clustering an effect associated with the region surrounding the hosts,
ideally one would like to measure its influence as closer as possible to the host.
In practice, this is done by measuring the amount of satellite candidates in
different annuli beyond our search radius \citep{Chen2006, Liu2011, Marmol-Queralto2012,
Ruiz2014}. We denote N$_{\rm C}$ as the number of 'satellite' galaxies placed in these
annuli which fulfil our selection criteria for each morphological host type. N$_{\rm C}$ is a
measurement of the background contamination plus the excess over this background
caused by the clustering. This method has the disadvantage, compared to the simulations
that we have conducted above, that is statistically more uncertain since N$_{\rm C}$
can be measured only around our massive galaxies and their number is relatively small.

To quantify the effect of the clustering, we count the number N$_{\rm C}$ of satellites 
observed in annuli between 500 and 600 kpc which fulfil our selection criteria. 
The size of each annuli has the same area that our main 
exploration area around the hosts [i.e. ${\rm \pi}$ (300 kpc)$^{2}$]. As it was
done  in the above section, we subtract N$_{\rm C}$ to the number of observed satellites  
N$_{\rm Obs}$ to correct for the statistic excess given by the clustering
in the sample of observed satellites. In other words:

\begin{equation}\label{formula_frac_clu}
N_{\rm Sat,C} = N_{\rm Obs} - N_{\rm C}
\end{equation}

The radial range 500-600 kpc for determining the clustering is a compromise among having a
local measurement of the environment around our massive host galaxies
but being far away enough such as the probability of finding a gravitationally
bounded satellite to our targeted galaxy will be low. The projected radial 
distance of 500 kpc is chosen following many works in the literature \citep[e.g.][]{Sales2004,
Sales2005,Chen2006,Bailin2008,Wang2012,Ruiz2014} which have used only galaxies with radial
distances lower than 500 kpc to define their sample of truly (i.e. bounded) satellite galaxies.

\section{Results}\label{results}

\subsection{Cumulative number of satellites per galaxy host}\label{sub:satcandfound}

\begin{figure*}
\includegraphics[width=1.0\textwidth,clip=true]{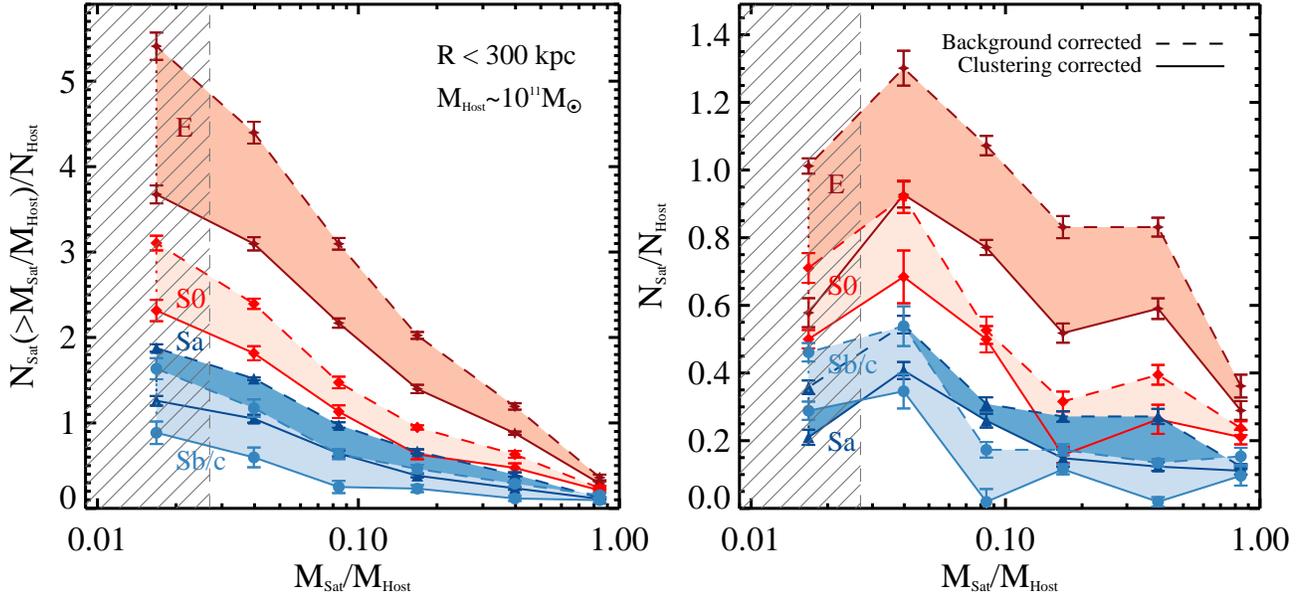}

\caption{
Left-hand panel: cumulative number of satellites per galaxy host for each morphological
        type versus the stellar mass ratio satellite-host down to 1:100. The dark red lines
        correspond to centrals E type, red to S0 type, dark blue to Sa type and soft blue to Sb/c type.
        The coloured areas represent the space between the abundance 
        estimated after applying background and clustering corrections. The dashed 
        lines correspond to the background correction and the continuous line to the
        clustering. The striped grey region indicates our more conservative measure
        of the completeness ($\sim$1:40) -- see Section ~\ref{sec:compspeccat}. 
        Right-hand panel: differential number of satellites per galaxy host for each of
        the morphological type versus the stellar mass ratio satellite--host. The 
        colour code associated with the lines is the same that in the left-hand panel. 
} 

\label{abundances} 
\end{figure*}

The abundance of satellites is quantified using the number of satellites per 
number of massive galaxies N$_{\rm Sat}$/N$_{\rm Host}$ down to a mass ratio
satellite-host of 1:100. The results are shown in the left-hand panel of the Fig.
~\ref{abundances} and Table \ref{table_abundances}. In addition to explore our 
satellites in a search radius of 300 kpc, we repeat the same exercise using a 100 kpc
radius to compare with previous results in the literature. 

\begin{table*}

\caption{
 Cumulative number of satellites per host galaxy using the background
 ($N_{\rm Sat,S}/N_{\rm Host}$) and clustering corrections ($N_{\rm Sat,C}/N_{\rm Host}$) within
 a radial distance of 300 kpc and down to a mass ratio satellite-host 1:100. 
 $N_{\rm Obs}/N_{\rm Host}$ is the observed number of satellites per host.
 $N_{\rm S}/N_{\rm Host}$ is the number of satellites per 
 host in our background simulation and $N_{\rm C}/N_{\rm Host}$ the number of
 satellites per host within 500-600 kpc in our clustering analysis.
 The background and clustering corrections are estimated for each morphological type as it is shown in the table.
     }
\begin{tabular}{lccccc}
\hline\hline
$M_{\rm Sat}/M_{\rm Host}$ & $N_{\rm Obs}/N_{\rm Host}$ & $N_{\rm S}/N_{\rm Host}$ & $N_{\rm C}/N_{\rm Host}$ & $N_{\rm Sat,S}/N_{\rm Host}$  & $N_{\rm Sat,C}/N_{\rm Host}$\\
\hline
E\\            
\hline
0.50-1.0&$      0.40\pm      0.03$&$     0.006\pm    0.0001$&$     0.080\pm     0.008$&$      0.36\pm      0.03$&$      0.29\pm      0.03$\\
0.20-1.0&$      1.23\pm      0.04$&$     0.021\pm    0.0002$&$     0.325\pm     0.022$&$      1.19\pm      0.04$&$      0.88\pm      0.02$\\
0.10-1.0&$      1.94\pm      0.04$&$     0.038\pm    0.0003$&$     0.647\pm     0.046$&$      2.02\pm      0.04$&$      1.40\pm      0.05$\\
0.05-1.0&$      3.14\pm      0.08$&$     0.056\pm    0.0004$&$     0.985\pm     0.061$&$      3.10\pm      0.07$&$      2.17\pm      0.05$\\
0.02-1.0&$      4.44\pm      0.11$&$     0.085\pm    0.0008$&$     1.373\pm     0.083$&$      4.40\pm      0.13$&$      3.10\pm      0.08$\\
0.01-1.0&$      5.51\pm      0.16$&$     0.113\pm    0.0010$&$     1.840\pm     0.105$&$      5.41\pm      0.16$&$      3.67\pm      0.11$\\
\hline
S0\\
\hline
0.50-1.0&$      0.23\pm      0.02$&$     0.005\pm    0.0001$&$     0.032\pm     0.010$&$      0.24\pm      0.02$&$      0.21\pm      0.02$\\
0.20-1.0&$      0.64\pm      0.04$&$     0.020\pm    0.0003$&$     0.168\pm     0.019$&$      0.63\pm      0.04$&$      0.47\pm      0.05$\\
0.10-1.0&$      0.90\pm      0.02$&$     0.037\pm    0.0006$&$     0.329\pm     0.019$&$      0.95\pm      0.02$&$      0.63\pm      0.06$\\
0.05-1.0&$      1.40\pm      0.07$&$     0.056\pm    0.0008$&$     0.346\pm     0.020$&$      1.47\pm      0.07$&$      1.13\pm      0.07$\\
0.02-1.0&$      2.35\pm      0.06$&$     0.084\pm    0.0012$&$     0.629\pm     0.040$&$      2.39\pm      0.06$&$      1.82\pm      0.08$\\
0.01-1.0&$      2.99\pm      0.09$&$     0.113\pm    0.0016$&$     0.804\pm     0.043$&$      3.11\pm      0.09$&$      2.32\pm      0.12$\\
\hline
Sa\\
\hline
0.50-1.0&$      0.11\pm      0.01$&$     0.006\pm    0.0001$&$     0.017\pm     0.003$&$      0.12\pm      0.01$&$      0.11\pm      0.01$\\
0.20-1.0&$      0.41\pm      0.03$&$     0.022\pm    0.0002$&$     0.164\pm     0.014$&$      0.40\pm      0.03$&$      0.23\pm      0.02$\\
0.10-1.0&$      0.67\pm      0.03$&$     0.039\pm    0.0003$&$     0.301\pm     0.033$&$      0.67\pm      0.03$&$      0.38\pm      0.02$\\
0.05-1.0&$      0.97\pm      0.03$&$     0.057\pm    0.0004$&$     0.349\pm     0.037$&$      0.98\pm      0.03$&$      0.64\pm      0.03$\\
0.02-1.0&$      1.45\pm      0.02$&$     0.086\pm    0.0006$&$     0.492\pm     0.053$&$      1.52\pm      0.02$&$      1.05\pm      0.05$\\
0.01-1.0&$      1.89\pm      0.04$&$     0.115\pm    0.0008$&$     0.662\pm     0.056$&$      1.88\pm      0.04$&$      1.26\pm      0.06$\\
\hline
Sb/c\\
\hline
0.50-1.0&$      0.13\pm      0.03$&$     0.006\pm    0.0001$&$     0.041\pm     0.008$&$      0.15\pm      0.03$&$      0.10\pm      0.03$\\
0.20-1.0&$      0.26\pm      0.03$&$     0.022\pm    0.0003$&$     0.158\pm     0.015$&$      0.29\pm      0.03$&$      0.12\pm      0.03$\\
0.10-1.0&$      0.42\pm      0.05$&$     0.039\pm    0.0005$&$     0.218\pm     0.019$&$      0.46\pm      0.05$&$      0.23\pm      0.04$\\
0.05-1.0&$      0.55\pm      0.06$&$     0.057\pm    0.0007$&$     0.368\pm     0.019$&$      0.63\pm      0.06$&$      0.25\pm      0.07$\\
0.02-1.0&$      1.02\pm      0.10$&$     0.086\pm    0.0009$&$     0.555\pm     0.035$&$      1.17\pm      0.10$&$      0.60\pm      0.12$\\
0.01-1.0&$      1.45\pm      0.12$&$     0.114\pm    0.0010$&$     0.655\pm     0.023$&$      1.63\pm      0.12$&$      0.88\pm      0.13$\\
\hline\hline
\end{tabular}
\label{table_abundances}
\end{table*}

The  N$_{\rm Sat}$/N$_{\rm Host}$ values have been corrected from contaminants by subtracting to 
$N_{\rm Obs}$ the quantities $N_{\rm S}$ and $N_{\rm C}$ found in the background simulation
and in our estimates of clustering, respectively. The uncertainties of $N_{\rm Obs}$ and $N_{\rm C}$
were estimated from bootstrap resamplings of host sample sets. 
As we can see in Table \ref{table_abundances}, the background contamination is very low  since our
work uses spectroscopic redshifts. According 
to that table, the maximum contamination expected by the background is $\sim$0.1 satellites per 
galaxy host. The ratios N$_{\rm S}/$N$_{\rm Host}$ and  N$_{\rm C}$/N$_{\rm Host}$ in Table
 \ref{table_abundances} show that the background contamination is, as expected, quite independent
on the morphological type. However, if we compare the typical number of fake satellites due to 
clustering ($N_{\rm C}/N_{\rm Host}$), we see significant differences.
By far, the host galaxies that are more likely affected by contaminants are the E-type. 
On one hand, we have to take into account that the massive Sb/c-types are not usually 
expected to be into the cluster's core regions. Thus, if we compare both $N_{\rm S}/N_{\rm Host}$ and
$N_{\rm C}/N_{\rm Host}$ around the massive Sb/c-types, we find a factor of 6 larger
contamination due to clustering than due to background. In contrast, this ratio increases 
dramatically up to $\sim$16 times for the elliptical hosts. S0 and Sa-types show a ratio
of $\sim$7-8 and $\sim$6-7, respectively, a factor similar to the one found around 
Sb/c-types. It is remarkable that the density of objects at assessing the clustering and
the background contaminant around the Sb/c-types, although close, are not similar. All
this indicates that these massive galaxies are immersed in an environment similar to the 
samples of S0 and Sa and therefore, they are not completely isolated.

Focusing on the satellite abundances, we find that the number of satellites
per galaxy host is 1.4-2.0 when we explore satellites down to a mass ratio 1:10 around
massive elliptical galaxies. If we increase that range of mass ratio down to 1:100,
we find 3.7-5.4. The other morphological types show less number of 
satellites than in the E-host case. The more extreme case is found at comparing with 
the massive Sb/c type, their number of satellites per galaxy host grows between 
0.23-0.46 and 0.88-1.63 from 1:10 to 1:100. Consequently, the massive
E hosts have $\sim$5 times more satellites than the massive Sb/c down to 1:10. A 
difference which declines up to 3-4 times in the case 1:100. At comparing
N$_{\rm Sat}$/N$_{\rm Host}$ down to 1:10 with the other samples of galaxies, we find that
the S0 and Sa types host $\sim$2 and 3 times less satellites, respectively than the 
E types. At extending our search down to 1:100, that difference among the S0--Sa types
and ellipticals barely change. When we restrict our satellite search up to only 100 kpc,
the difference in the number of satellites among the different morphological types 
remains similar. As expected, the total number of satellites decreases when comparing 
a search radius of 300 kpc to one of 100 kpc. The decreasing factor is 2.5, 2.9, 2.2 and
3.5 for the E, S0, Sa and Sb/c types, respectively. 

\subsection{Differential number of satellites per galaxy host}\label{sub:satcandfoundif}

We have also explored different intervals of the mass ratio satellite--host.  
The intervals are defined as 1:1-1:2, 1:2-1:5, 1:5-1:10, 1:10-1:20, 1:20-1:50,
1:50-1:100 and the number of satellites per host is illustrated in the
right-hand panel of the Fig.~\ref{abundances} and Table~\ref{table_abundances_dif}. 
This `differential' test allows us to compare how is distributed the population of
satellites respect to their stellar mass. 

As the results presented above, the dependence of the abundance of satellites with 
the morphology of the host is also appreciable. In general, the massive ellipticals have 
more satellites, followed by S0, Sa and Sb/c types.

As shown in Table ~\ref{table_abundances_dif}, the number of fake satellites
per host due to the background $N_{\rm S}/N_{\rm Host}$ remains independently on the 
morphological type whereas the contamination due to the clustering 
$N_{\rm C}/N_{\rm Host}$ is prominently larger in the elliptical case and very 
similar among the types S0, Sa and Sb/c. 

\begin{table*}
\centering
\caption{
 Differential number of satellites per host galaxy using the background
 ($N_{\rm Sat,S}/N_{\rm Host}$) and clustering corrections ($N_{\rm Sat,C}/N_{\rm Host}$) within
 a radial distance of 300 kpc and down to a mass ratio satellite--host 1:100. 
 $N_{\rm Obs}/N_{\rm Host}$ is the observed number of satellites per host.
 $N_{\rm S}/N_{\rm Host}$ is the number of satellites per 
 host in our background simulation and $N_{\rm C}/N_{\rm Host}$ the number of
 satellites per host within 500-600 kpc in our clustering analysis.
 The background and clustering corrections are derived independently for each morphological type as it is shown in the table.
       }
\begin{tabular}{lccccc}
\hline\hline
$M_{\rm Sat}/M_{\rm Host}$ & $N_{\rm Obs}/N_{\rm Host}$ & $N_{\rm S}/N_{\rm Host}$ & $N_{\rm C}/N_{\rm Host}$ & $N_{\rm Sat,S}/N_{\rm Host}$ & $N_{\rm Sat,C}/N_{\rm Host}$\\
\hline
E\\            
\hline
0.50-1.00&$      0.40\pm      0.03$&$     0.006\pm    0.0001$&$     0.080\pm     0.008$&$      0.36\pm      0.03$&$      0.29\pm      0.03$\\
0.20-0.50&$      0.83\pm      0.03$&$     0.015\pm    0.0002$&$     0.251\pm     0.022$&$      0.83\pm      0.03$&$      0.59\pm      0.03$\\
0.10-0.20&$      0.78\pm      0.03$&$     0.017\pm    0.0002$&$     0.322\pm     0.032$&$      0.83\pm      0.03$&$      0.52\pm      0.03$\\
0.05-0.10&$      1.13\pm      0.03$&$     0.018\pm    0.0002$&$     0.338\pm     0.020$&$      1.07\pm      0.03$&$      0.77\pm      0.02$\\
0.02-0.05&$      1.36\pm      0.05$&$     0.029\pm    0.0003$&$     0.389\pm     0.026$&$      1.30\pm      0.05$&$      0.93\pm      0.04$\\
0.01-0.02&$      0.94\pm      0.02$&$     0.028\pm    0.0003$&$     0.467\pm     0.027$&$      1.01\pm      0.02$&$      0.58\pm      0.04$\\
\hline
S0\\
\hline
0.50-1.00&$      0.23\pm      0.02$&$     0.005\pm    0.0001$&$     0.032\pm     0.010$&$      0.24\pm      0.02$&$      0.21\pm      0.02$\\
0.20-0.50&$      0.41\pm      0.03$&$     0.014\pm    0.0002$&$     0.137\pm     0.013$&$      0.39\pm      0.03$&$      0.26\pm      0.04$\\
0.10-0.20&$      0.30\pm      0.03$&$     0.017\pm    0.0003$&$     0.174\pm     0.018$&$      0.32\pm      0.03$&$      0.16\pm      0.02$\\
0.05-0.10&$      0.46\pm      0.04$&$     0.019\pm    0.0003$&$     0.005\pm     0.003$&$      0.53\pm      0.04$&$      0.50\pm      0.04$\\
0.02-0.05&$      0.91\pm      0.05$&$     0.028\pm    0.0004$&$     0.255\pm     0.021$&$      0.92\pm      0.05$&$      0.68\pm      0.08$\\
0.01-0.02&$      0.63\pm      0.04$&$     0.029\pm    0.0004$&$     0.141\pm     0.016$&$      0.71\pm      0.04$&$      0.50\pm      0.03$\\
\hline
Sa\\
\hline
0.50-1.00&$      0.11\pm      0.01$&$     0.006\pm    0.0001$&$     0.017\pm     0.003$&$      0.12\pm      0.01$&$      0.11\pm      0.01$\\
0.20-0.50&$      0.27\pm      0.02$&$     0.016\pm    0.0002$&$     0.147\pm     0.014$&$      0.27\pm      0.02$&$      0.12\pm      0.01$\\
0.10-0.20&$      0.29\pm      0.02$&$     0.017\pm    0.0002$&$     0.137\pm     0.023$&$      0.27\pm      0.02$&$      0.15\pm      0.03$\\
0.05-0.10&$      0.30\pm      0.02$&$     0.018\pm    0.0001$&$     0.048\pm     0.009$&$      0.31\pm      0.02$&$      0.26\pm      0.02$\\
0.02-0.05&$      0.57\pm      0.03$&$     0.029\pm    0.0002$&$     0.143\pm     0.017$&$      0.54\pm      0.03$&$      0.41\pm      0.03$\\
0.01-0.02&$      0.40\pm      0.02$&$     0.029\pm    0.0002$&$     0.156\pm     0.014$&$      0.36\pm      0.02$&$      0.21\pm      0.02$\\
\hline
Sb/c\\
\hline
0.50-1.00&$      0.13\pm      0.03$&$     0.006\pm    0.0001$&$     0.041\pm     0.008$&$      0.15\pm      0.03$&$      0.10\pm      0.03$\\
0.20-0.50&$      0.13\pm      0.01$&$     0.016\pm    0.0002$&$     0.125\pm     0.014$&$      0.13\pm      0.01$&$      0.02\pm      0.01$\\
0.10-0.20&$      0.14\pm      0.02$&$     0.017\pm    0.0003$&$     0.046\pm     0.009$&$      0.17\pm      0.02$&$      0.12\pm      0.01$\\
0.05-0.10&$      0.15\pm      0.02$&$     0.019\pm    0.0002$&$     0.150\pm     0.015$&$      0.17\pm      0.02$&$      0.02\pm      0.04$\\
0.02-0.05&$      0.51\pm      0.06$&$     0.028\pm    0.0002$&$     0.186\pm     0.020$&$      0.54\pm      0.06$&$      0.35\pm      0.05$\\
0.01-0.02&$      0.43\pm      0.03$&$     0.028\pm    0.0002$&$     0.153\pm     0.016$&$      0.46\pm      0.03$&$      0.29\pm      0.03$\\
\hline\hline
\end{tabular}
\label{table_abundances_dif}
\end{table*}

\subsection{The amount of mass surrounding the galaxy hosts}\label{distrib_mass}

\begin{figure}
\includegraphics[width=1.0\columnwidth,clip=true]{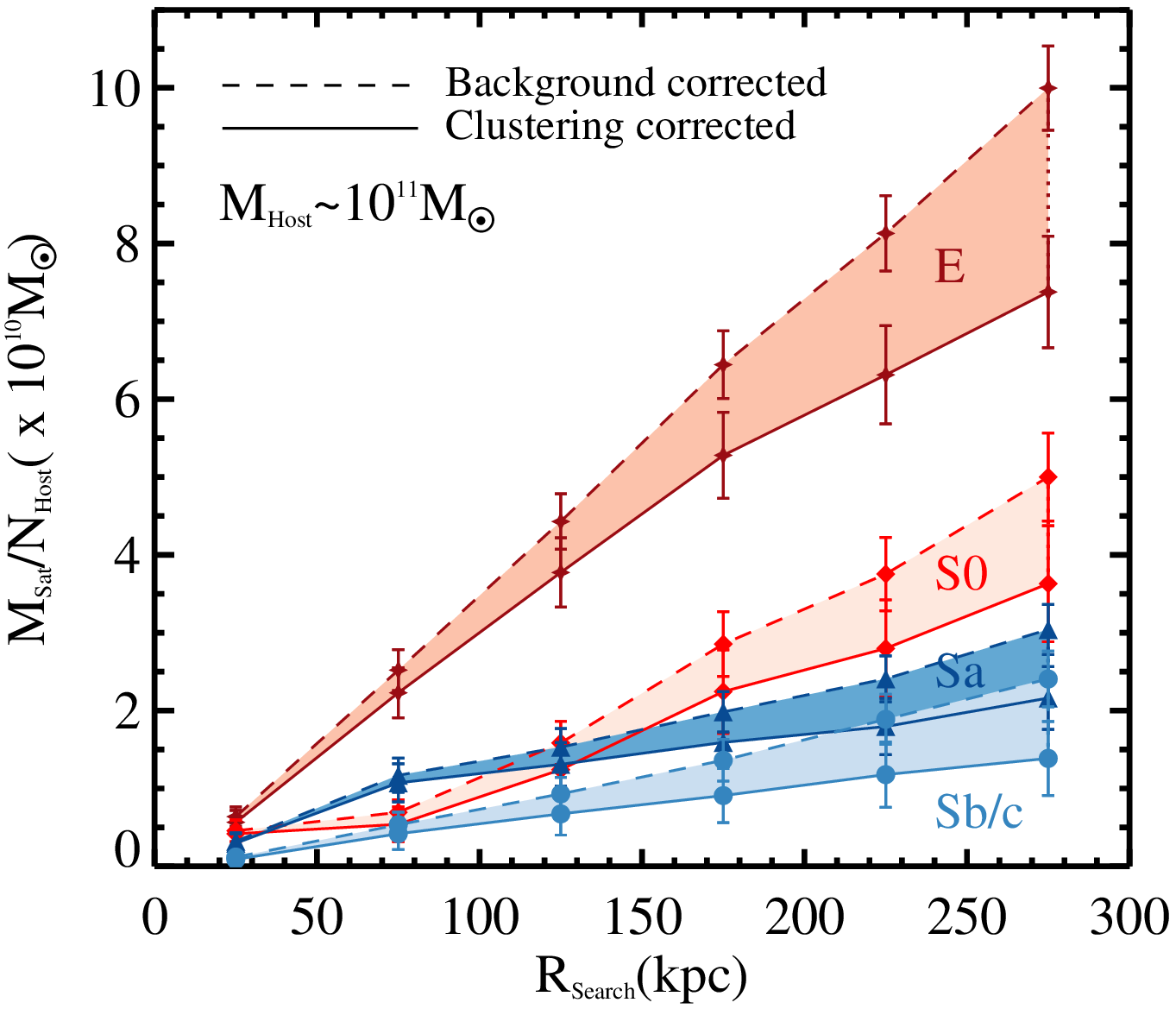}
\caption{
Cumulative stellar mass enclosed by satellites down to 1:100 
as a function of the projected radial distance up to 300 kpc. The background 
contaminant correction, as well as the clustering correction, is applied for 
all the morphological types. The dark red lines correspond to centrals E, red S0, 
dark blue Sa and soft blue Sb/c type. 
} 
\label{cumulativesatellitemass} 
\end{figure}

We have studied the amount of stellar mass accumulated by the satellites 
around our samples of massive galaxies. This is estimated down to 1:100 and
as a function of their projected distances to the central galaxy.
This cumulative stellar mass of the satellites is measured summing the 
stellar mass of all the satellites down to 1:100 in each interval of the 
search radius up to 300 kpc and then, subtracting the amount of stellar mass
in the fake satellites from the background simulations. To apply the clustering
correction on the amount of stellar mass enclosed by the satellites, we estimate
the amount of mass in 'satellites' in the interval 500-600 kpc using an annuli
with the same projected area than the ones used to study this quantity (i.e. from
${\rm \pi}$ 50$^{2}$ to ${\rm \pi}$ 300$^{2}$). The results are illustrated in 
Fig.~\ref{cumulativesatellitemass}. 
 
E-type massive galaxies are surrounded by a factor of 2-5 more 
stellar mass respect to the rest of morphological types. If we repeat this
exercise using only satellites up to 100 kpc, the stellar mass accumulated for 
lenticulars and spirals is similar, however, it is still a factor of 2-6 lower than the
mass collected by ellipticals. Among the early types, the ratio of mass 
surrounding ellipticals to the mass surrounding lenticulars is typically placed
around a factor of 2 larger. The spiral types show a more gradual accumulation
of stellar mass with radius. It is worth noting that the amount of stellar mass
enclosed by the satellites of massive ellipticals is almost as large as the mass of the
host galaxies (i.e. $\sim$10$^{11}$M$_{\sun}$). The error bars in 
Fig.~\ref{cumulativesatellitemass} are estimated 
using the contribution of Poisson errors based on the number of observed
satellites and the standard deviation of the average number of fake satellites 
found in the simulations.

\subsection{The merging channel of massive galaxies}\label{sub:merging_channel}

Using the previous distribution of satellites, we can speculate
about the stellar mass which could be potentially transferred to the hosts due to satellite
infall. Under the assumption that eventually, all the satellites surrounding our 
massive galaxies will infall into their massive hosts,  we can estimate which satellites
could, in the future,  contribute most to a potential mass growth of the host galaxy.
On what follows, we assume that all satellites, independently of their mass, will 
infall with the same speed on the central galaxy. Note, however, that this could be 
not necessary true, since it is theoretically expected that most massive
satellites will have shorter merging time-scale \citep[e.g.][]{Jiang2013}.

To probe the most likely merger channel what we have done is the following. We have added all
the stellar mass contained by the satellites within the intervals of stellar mass studied, then, 
we divide this quantity by the sum of the mass of all the host galaxies, such as it indicates the
following equation:

\begin{equation}
\Psi=\frac{\sum\limits_{i=1}^{N_{\rm Sat-bin}}M_{{\rm Sat-bin},i}}{\sum\limits_{j=1}^{N_{\rm Host}} M_{{\rm Host},j}} 
\end{equation}

\begin{figure*}
     \begin{center}
            \includegraphics[width=0.9\textwidth]{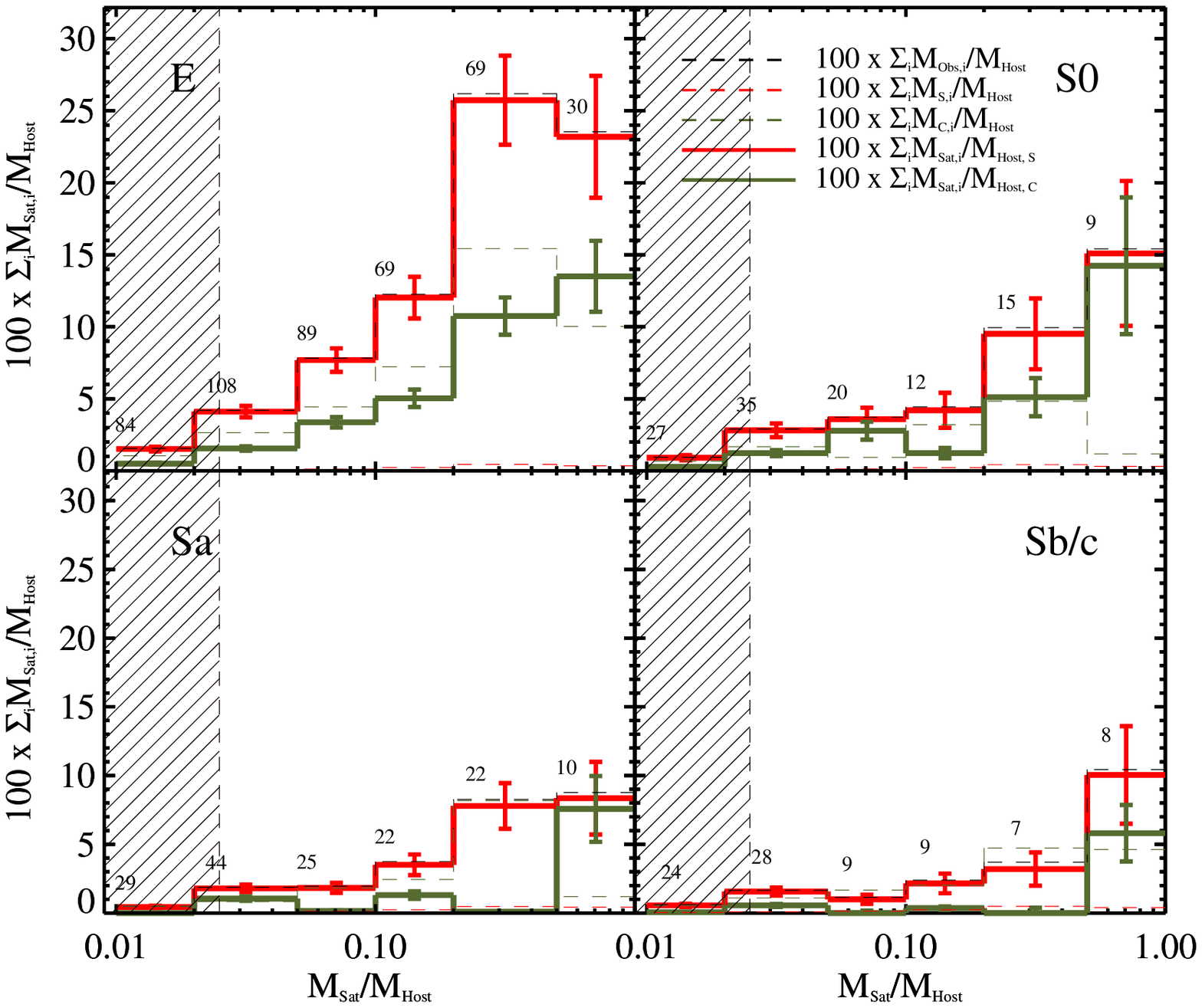}
\end{center}
\caption{%
The merging channel of the massive galaxies as a function of their morphology. The panels
show the contribution (in percentage) of the satellite mass enclosed in each mass
bin to the total mass confined by their hosts for each morphological type. 
The red solid line represents this quantity after correcting for the background contaminant
($\Psi_{\rm S}$),  and the green solid line after correcting by clustering ($\Psi_{\rm C}$). 
The black, red and green dashed lines show $\Psi_{\rm Obs}$,
$\Psi_{\rm Sim}$ and $\Psi_{\rm Clu}$ found in the  observations, background simulations and
clustering, respectively. The numbers over each bin correspond to the number of
observed satellites within each mass interval. The vertical dashed area illustrates 
the region where the satellite incompleteness could play a role.
     }%
   \label{figure3}
\end{figure*}

\begin{table}
\centering
\caption{
The merging channel of local massive galaxies for each morphological type. 
The table shows the contribution (in per cent) of the stellar mass enclosed in each satellite mass 
bin to the total mass confined by their hosts. We show the observed fraction (in per cent)
$\Psi_{\rm Obs}$ as well as their values once corrected from background $\Psi_{\rm Sat,S}$
and clustering $\Psi_{\rm Sat,C}$.
}

\begin{tabular}{lccc}
\hline
$M_{\rm Sat}/M_{\rm Host}$ & $\Psi_{\rm Obs}$  &  $\Psi_{\rm Sat,S}$ & $\Psi_{\rm Sat,C}$\\
                           & (per cent)        & (per cent)          & (per cent)        \\
\hline
E \\
0.50-1.00&$     23.54\pm      4.30$&$     23.19\pm      4.23$&$     13.50\pm      2.47$\\
0.20-0.50&$     26.18\pm      3.15$&$     25.73\pm      3.10$&$     10.75\pm      1.29$\\
0.10-0.20&$     12.26\pm      1.48$&$     12.03\pm      1.45$&$      5.04\pm      0.61$\\
0.05-0.10&$      7.82\pm      0.83$&$      7.69\pm      0.82$&$      3.38\pm      0.36$\\
0.02-0.05&$      4.21\pm      0.40$&$      4.12\pm      0.40$&$      1.56\pm      0.15$\\
0.01-0.02&$      1.56\pm      0.17$&$      1.52\pm      0.17$&$      0.50\pm      0.06$\\
Total&$     75.58\pm     10.33$&$     74.27\pm     10.16$&$     34.73\pm      4.93$\\
S0  \\
0.50-1.00&$     15.42\pm      5.14$&$     15.09\pm      5.03$&$     14.23\pm      4.74$\\
0.20-0.50&$      9.95\pm      2.57$&$      9.51\pm      2.46$&$      5.12\pm      1.32$\\
0.10-0.20&$      4.44\pm      1.28$&$      4.21\pm      1.21$&$      1.23\pm      0.36$\\
0.05-0.10&$      3.73\pm      0.83$&$      3.59\pm      0.80$&$      2.79\pm      0.62$\\
0.02-0.05&$      2.91\pm      0.49$&$      2.82\pm      0.48$&$      1.23\pm      0.21$\\
0.01-0.02&$      0.96\pm      0.18$&$      0.92\pm      0.18$&$      0.28\pm      0.05$\\
Total&$     37.40\pm     10.50$&$     36.14\pm     10.16$&$     24.89\pm      7.31$\\

Sa  \\
0.50-1.00&$      8.76\pm      2.77$&$      8.35\pm      2.64$&$      7.57\pm      2.39$\\
0.20-0.50&$      8.26\pm      1.76$&$      7.79\pm      1.66$&$      0.09\pm      0.02$\\
0.10-0.20&$      3.75\pm      0.80$&$      3.51\pm      0.75$&$      1.30\pm      0.28$\\
0.05-0.10&$      1.96\pm      0.39$&$      1.83\pm      0.37$&$      0.15\pm      0.03$\\
0.02-0.05&$      1.88\pm      0.28$&$      1.79\pm      0.27$&$      1.05\pm      0.16$\\
0.01-0.02&$      0.48\pm      0.09$&$      0.44\pm      0.08$&$      0.00\pm      0.01$\\
Total&$     25.09\pm      6.10$&$     23.70\pm      5.77$&$     10.17\pm      2.88$\\
Sb/c  \\
0.50-1.00&$     10.43\pm      3.69$&$     10.04\pm      3.55$&$      5.81\pm      2.05$\\
0.20-0.50&$      3.69\pm      1.40$&$      3.20\pm      1.21$&$      0.00\pm      0.39$\\
0.10-0.20&$      2.38\pm      0.79$&$      2.15\pm      0.72$&$      0.36\pm      0.12$\\
0.05-0.10&$      1.14\pm      0.38$&$      1.00\pm      0.33$&$      0.00\pm      0.18$\\
0.02-0.05&$      1.65\pm      0.31$&$      1.56\pm      0.30$&$      0.56\pm      0.11$\\
0.01-0.02&$      0.60\pm      0.12$&$      0.56\pm      0.11$&$      0.06\pm      0.01$\\
Total&$     19.89\pm      6.69$&$     18.51\pm      6.22$&$      6.78\pm      2.86$\\
\hline
\end{tabular}
\label{merging_channel_tab}
\end{table}

The sum of all the mass in the host galaxies is a fixed quantity for our samples of hosts 
and their values are  $\sum_{j=1}^{N_{\rm Host}} M_{{\rm Host},j}$=(11.2, 5.26, 10.4,
6.76)$\times$10$^{12}$M$_{\sun}$ for (E, S0, Sa, Sb/c) types, respectively. These numbers
correspond to the following typical masses per galaxy host:
$\sum_{j=1}^{N_{\rm Host}} M_{{\rm Host},j}/N_{\rm Host}$=(1.3, 1.4, 1.3, 1.3)
$\times$10$^{11}$M$_{\sun}$.

Our results are illustrated in Fig.~\ref{figure3} and
Table ~\ref{merging_channel_tab} for each host sample, once corrected by the effects of
background and clustering. Fig.~\ref{figure3} highlights the different way the stellar
mass around the massive galaxies builds up as a function of the morphological type.
The merger channel is mainly produced by satellites down to 1:10 for all the host samples. The
average total amount of stellar mass contained in the satellite population down to 1:10 compared 
to the total amount of stellar mass in the hosts to the different morphological types
studied is 45.4$\pm$6.6, 24.7$\pm$7.5, 14.3$\pm$3.8, 10.8$\pm$3.7 per cent for E, S0, Sa, Sb/c types,
respectively. Down to 1:100 these values increase up to 54.5$\pm$15.1, 
30.5$\pm$8.7, 16.9$\pm$4.3, 12.6$\pm$4.5 per cent of the total amount of mass contained in 
their hosts. These numbers indicate that the contribution of the low mass population of satellites
1:10-1:100 to the host mass is not significant compared to the mass ratio 1:1-1:10. 
We have assumed Poisson errors to estimate our error bars in the Fig.~\ref{figure3}.

At limiting our exploration up to 100 kpc, the above results decrease notably. The satellites
more massive than 1:10 are again the dominant mass contributor. The percentage
which reflects that contribution with respect to the total mass of the hosts is roughly
14.6$\pm$3.2, 2.4$\pm$1.3, 7.9$\pm$1.0, 3.0$\pm$2.1 per cent for E, S0, Sa, Sb/c types, respectively,
corresponding to the mean of background-clustering results once applied the corrections.
Down to 1:100, these percentages increase up to 17.2$\pm$3.9, 4.5$\pm$2.1,
8.7$\pm$2.8, 3.6$\pm$2.4.

As we pointed out before, if the satellites eventually infall into the host galaxies, 
the merger channel will be largely dominated by satellites with a mass ratio below 1:10 
for all morphological types. For satellites up to 300 kpc (100 kpc), this corresponds to 67.2,
68.4, 88.1 and 85.7 (85, 57, 91 and 83) per cent of the total mass enclosed by the
satellites within that radial distance. The contribution of the most massive satellites 
is particularly important for spiral types, and slightly weaker for early types. Down to 1:10, 
the distribution of stellar mass of the satellites reveals that the massive E galaxies
could have their main contributor between the relative masses 1:5-1:1 whereas the main contributor 
of the S0, Sa and Sb/c types is the most massive satellites (1:2-1:1). It is also worth 
noticing that elliptical galaxies can have up to $\sim$4 times more mass in satellites 
of 1:10-1:100 compared to Sb/c types. 

If the theoretical expectations are correct, and the merger time-scales are shorter
for the most massive satellites, this mass growth due to the larger satellites will be even
more important than the result shown in Fig.~\ref{figure3}.

\section{Discussion}\label{sec:discussion}

The results presented in this paper attempt to establish a robust $z\sim$0 reference for the study 
of the evolution of the abundance of satellites around massive galaxies with cosmic time and 
their dependence with host morphology. In addition, it extends the results obtained in
our previous work, \citet{Ruiz2014}, in which we studied the abundance of satellites around E-type hosts using 
the SDSS Data Release 7 (DR7). In this paper, we have completed that results, using a deeper
spectroscopic sample and studying the satellite abundances also according to their host morphology. 

We have found that the abundance of satellites turns out to be significantly dependent 
on the morphology of the hosts for galaxies with similar stellar masses. The abundance of satellites 
is much higher around elliptical galaxies. The fact that Sb/c galaxies have fewer satellites
around them probably helps to explain how they have maintained their disc-like structure during their lifetime. 
On the contrary, the large number of satellites around massive ellipticals could help to clarify
the characteristic large envelopes surrounding these objects that are thought to be connected 
with intensive accretion. 

Several works have recently estimated the number of satellites in different redshift intervals
around massive galaxies both observationally and theoretically
\citep[see e.g.][]{Chen2008,Moster2010, {QuanGuo2011,QuanGuo2013},Lares2011, Liu2011,Nierenberg2012,
Quilis2012, Tal2012, Wang2012, Wang2013}. The observational studies agree
within the measurement uncertainties. However, the satellites in the  simulations are, 
in general, more abundant.
Some of these previous studies have shown that the number of satellites around massive galaxies 
of a given mass ratio has remained constant since at least $z\sim$2 ~\citep{Jackson2010,  {Nierenberg2011,Nierenberg2013},Man2012, {Marmol-Queralto2012,Marmol-Queralto2013},
Newman2012, Quilis2012, Huertas2013, Ferreras2014}. 
In this paper, less  affected by incompleteness at low stellar mass than 
previous works, we readdress this question and explore how the theoretical expectations 
compare to the observational data in the local Universe. We do this for our four different 
morphological types. 

\subsection{Abundance of satellites: comparison with previous works}

\subsubsection{Observations}\label{pos}

In order to check the agreement of our results with the literature, 
we do a direct comparison with the abundance of satellites found by~\citet{Wang2012}. 
In their work, the authors studied the number of satellites around a large sample of 
isolated massive galaxies to probe the $\Lambda$CDM scenario.
In their paper, they segregated the massive galaxies not using visual morphology
but colours, and estimated the abundance of satellites according to the stellar mass of the
satellites instead of the mass ratio satellite--host.

To make a direct comparison with~\citet{Wang2012}, we also estimate the  abundance of
satellites according to the stellar mass instead the mass ratio satellite-host such
as~\citet{Wang2012} did. Given that the average stellar mass of our host sample is 
log(M$_{\star}$/M$_{\sun}$)$\sim$11.1, we have compared our numbers with the average value
they get combining their results from the host mass bins
10.8$<$log(M$_{\star}$/M$_{\sun}$)$<$11.1 and 11.1$<$log(M$_{\star}$/M$_{\sun}$)$<$11.4
(their green and blue lines, respectively, in their figs 7 and 8). We confront our
results for E and S0 types with the ones they found for their red hosts. Additionally, we
explore the difference between the abundance of satellites in their blue hosts with our
sample of disc galaxies (Sa and Sb/c types).  Thus, we obtain, for satellites with
log(M$_{\star}$/M$_{\sun}$)=10, $N_{\rm Sat}/N_{\rm Host}$= 0.31-0.44 (ours) and
0.14-0.19  (Wang \& White) and, for  satellites with log(M$_{\star}$/M$_{\sun}$)=9.0, 
$N_{\rm Sat}/N_{\rm Host}$= 0.50-0.65 (ours) and 0.44-0.55 (Wang \& White).  When we
compare our disc types with their blue massive galaxies we find,  for satellites with
log(M$_{\star}$/M$_{\sun}$)=10,  $N_{\rm Sat}/N_{\rm Host}$= 0.05-0.1 (ours) and 0.07-0.12
(Wang \& White) and, for satellites with log(M$_{\star}$/M$_{\sun}$)=9.0, $N_{\rm
Sat}/N_{\rm Host}$ =0.22-0.29 (ours) and 0.08-0.15 (Wang \& White). Our results are 
significantly larger at comparing their red primaries  with our early-type galaxies for
log(M$_{\star}$/M$_{\sun}$)=10 and  closer for less massive satellites
(log(M$_{\star}$/M$_{\sun}$)=9). This trend is not reproduced in the comparison of blue
primaries versus late-type galaxies where we find a closer agreement. 

The difference with \citet{Wang2012}  is even larger if we consider a more extreme
comparison:  our massive elliptical galaxies versus their red primaries. In this case, we
obtain for satellites with log(M$_{\star}$/M$_{\sun}$)=10, $N_{\rm Sat}/N_{\rm Host}$=
0.41-0.61 (ours) and 0.14-0.19  (Wang \& White) and, for  satellites with
log(M$_{\star}$/M$_{\sun}$)=9.0,  $N_{\rm Sat}/N_{\rm Host}$= 0.53-0.78 (ours) and
0.44-0.55 (Wang \& White). This highlights the importance of morphology when we study the
number of satellites. Several reasons can contribute to the differences found at comparing the 
red massive galaxies of \citet{Wang2012} with our early-type galaxies. 
The most important is probably the environment 
where the samples are immersed since their study focus only on isolated 
galaxies whereas we use all the galaxies. However, the abundance
of satellites is some closer at extending the comparison to the poorer mass 
satellites. This is likely due to our incompleteness around 10$^{9}$M$_{\sun}$. 
In contrast, this effect is not so evident at comparing blue-massive versus
our late-type galaxies since these galaxies more likely live in less crowded 
environment. 

From the above results we highlight two different characteristics. First, 
the isolation criteria used in previous works probably biases the global 
distribution of satellites around the massive galaxies. And second, 
the determination of abundance of satellites using morphological criteria
enhances the differences in the number of satellites compared those 
based on segregating the samples simply by colour. This implies a 
stronger connection between the galaxy morphology and their number 
of satellites larger than a potential connection between the satellites 
and the stellar population of their hosts. 

We can also compare our results with the recent work by ~\citet{Ruiz2014} on the 
satellite populations around massive elliptical galaxies in the local universe. 
In that work, we studied the abundance of satellites
within a radial distance of 100 kpc and selected
the host samples within the redshift range 0.02-0.065. We found, after applying 
the contamination corrections, that $N_{\rm Sat}/N_{\rm Host}$ was between 0.24 and 0.28 
for satellites up to 1:10 using a spectroscopic catalogue and $N_{\rm Sat}/N_{\rm Host}$  
0.84 and 0.98 when we extended our search up to 1:100 using a catalogue
of photometric redshifts (`photo-$z$'; SDSS DR7). In our new host 
sample of massive ellipticals, we obtain higher values 
for both ratios (0.36-0.43 for 1:10 and 1.12-1.31 for 1:100)
when we restrict our search radius down to R=100 kpc. The difference between the average
stellar mass of both samples is 0.1 dex, being the average of~\citet{Ruiz2014}'s sample larger. 
This difference then could be attributed to technical characteristics as the
effect of the 'tiling' of SDSS at different redshifts. In fact, in our previous work,
we estimated that we could approximately lose up to 25 per cent of 
our potential satellites due to fibre collisions, a quantity which could partially explain 
that difference in the case 1:10. Also, it could be playing a role the samples selected in both works. 
In fact, the selection of E galaxies done in \citet{Ruiz2014} is based on \citet{Nair2010} 
whereas the present classification is based on our own morphological analysis. Being
the galaxies in the present sample significantly closer than those in  \citet{Ruiz2014}, 
our ability to distinguish among E and S0 could be higher. In this sense, it worth stressing
that the abundance of satellites found around S0 massive galaxies for 1:10 and 1:100 is 0.12-0.15 
and 0.68-0.72, respectively. If we combine the E and S0 results, for 1:10, we get 
$\sim$0.24-0.29 and 0.9-1.05 for 1:100. This is also in good agreement with our previous 
results ~\citep{Ruiz2014}.

Leaving aside the number of satellites found per host, we can now draw our attention
to the fraction of massive galaxies with satellites found in other works at z=0, a 
parameter easily evaluable from our study. In this context, ~\citet{Liu2011}, 
exploring Milky Way-like galaxies, found that only 12 per cent of those objects
have at least a satellite within $R=$100 kpc down to 1:10 mass ratio 
(private communication). We have evaluated the fraction of massive host having
satellites down to 1:10 in our disc-like hosts. We find 10-12 per cent once
corrected from clustering and background contaminant. At segregating between the
Sa and Sb/c samples, that fraction is 0.15 and 0.07, respectively. Our results
are in good agreement with \citet{Liu2011}.

Also, we can make a comparison with ~\citet{Marmol-Queralto2012}. In this work, the authors
conducted a similar analysis to what we have done here but for galaxies at higher redshifts
(0.2$<$z$<$2). We compare our numbers with the galaxies they classified as spheroids in
their lower redshift range (0.2$<$z$<$0.75). They found that the fraction of massive galaxies
with satellites around the sample and down to 1:10 is 23-28 per cent. 
Our results for spheroids, E and S0 massive galaxies, are approximately 28 and 34 per cent, 
respectively. As we commented before, this comparison depends largely on the classification 
or grouping of the massive galaxies. A sample of spheroids with many S0 objects reduces 
the satellite abundance whereas a higher purity in the E-type sample increases it. We can 
also compare with the galaxies they classified as disc-like in the same redshift range, 
they find that the fraction of massive galaxies with at least a satellite was 5-9 per cent.
A closer but lower value to our disc-like 10-13 per cent. However, as we mentioned before, 
at splitting into the morphological types Sa and Sb/c (we obtained 15 and 7 per cent,
respectively) the agreement could depend again on the number of members of each 
morphological type.

Other works by \citet{Chen2008}, \citet{QuanGuo2011}, \citet{Nierenberg2012} or 
\citet{Kawinwanichakij2014} also searched for satellites around massive galaxies segregating
their samples by colours, early- and late-type morphologies or even quiescent and star-forming galaxies. 
Some of the results of these works can be compared with our results due to the similarity 
in the satellite search criteria. Nevertheless, these authors computed the potential satellites
as a function of the visual or $r$-band magnitude contrast satellite--host 
\citep{Chen2008,QuanGuo2011,Nierenberg2012} instead the mass ratio. Therefore, in order to 
compare with them, we also made our search for satellites using the $r$-band contrast magnitude 
between the satellite and its host.

In \citet{QuanGuo2011}, the authors studied the luminosity
function of satellite galaxies around isolated bright hosts (M$_r \sim$ -21.25) using SDSS galaxy 
samples. They scaled their search radius to 300 kpc and found that the average number of 
satellites per host was 0.1 for red hosts and 0.06 for blue hosts when they studied 
satellites for $\Delta$m$_v\sim$2.5 (see their fig.9). In addition, they split
their sample into early and late types, finding similar results. At comparing with their 
late or blue hosts, the results are far from our abundance of satellites using that $r$-band magnitude contrast,
we obtain N$_{\rm Sat}$/N$_{\rm Host}\sim$0.15-0.21. Also, when we compare with our results for early-type
galaxies, we obtain N$_{\rm Sat}$/N$_{\rm Host}\sim$0.41-0.60, an amount significantly higher than those 
of \citet{QuanGuo2011}. A possible explanation to this difference could be linked to their 
stellar mass distribution. To build their fig.9 they use host galaxies within the magnitude range
-20.75$<$M$_v<$-21.75 whereas our host sample is around -21.5$<$M$_r<$-22.5.

Interestingly, \citet{Chen2008} also studied the satellites around bright 
isolated galaxies, but unlike \citet{QuanGuo2011}, the authors chose their host sample
in the nearby Universe (z$<0.045$). That sample was somewhat brighter than 
\citet{QuanGuo2011}'s one (-20$<$M$_r<$-23) and they used a mildly shallower sample of satellites ($\Delta m_r<$ 2).
Under these conditions, they found an abundance of satellites per host of 0.3-0.4 and $\sim$1.0 for blue and
red hosts, respectively. This abundance of satellites found by \citet{Chen2008} is some lower than
our results for Sa and Sb/c types combined 0.42-0.60 for $\Delta m_r<$2  but close to our 
results around the Sb/c types (0.37-0.53) considering our clustering correction. 
Also, our abundance of satellites for (E, S0) hosts combined (1.31-1.79) is larger than 
which \citet{Chen2008} found. This likely due to the isolation criteria applied to 
select their host sample.  

As we have seen above, there are multiple factors which affect to comparison with other works. 
The usage of colours or morphology for selecting the hosts, the search criteria for satellites, 
the difference among the stellar mass distributions of confronted samples or the density of 
galaxies where these samples are immersed. Nonetheless and in general, we get a reasonable agreement 
with the literature when we compare galaxies with disc like morphologies, our Sa and Sb/c types. 
As we mentioned before, these massive galaxies are thought to live in less crowded environments
and therefore, their  abundance of satellites should not be susceptible to change too much at 
applying them an isolation criterion unlike what happens with early-type hosts and mainly with 
the massive ellipticals.  
The majority of works showed above have used host samples using an isolation criterion. 
This probably biases the global  distribution of satellites since it does not consider 
the excess of satellites linked to crowded regions in the real Universe. Precisely, in
these comparisons, our abundance of satellites around  early-type galaxies is, typically, 
larger, especially considering our sample of massive ellipticals.  In fact, this excess is 
softened due to the mixture of E and S0 galaxies \citep{Ruiz2014} since the 
S0 massive galaxies are typically in similar environments than the Sa types.

In that sense, a recent study by \citet{Guo2015} investigated the dependence of the luminosity function of 
the host galaxies as a function of inhabiting a filament or not using the SDSS data set. 
They found that the filamentary environment can increase the abundance of the brightest 
satellites by a factor of $\sim$2  compared with non-filament isolated galaxies. This can
help to explain the discrepancy  with other observational works which used isolation
 \citep[e.g.][]{Chen2008,QuanGuo2011,Wang2012} since many of our massive elliptical 
galaxies are expected to be in dense regions. 

\subsubsection{Theory}

\citet{Quilis2012} estimated, using the Millennium Simulations, the expected fraction of 
massive galaxies with satellites with a mass ratio down to 1:10 and down to 1:100 within
a sphere of $R=$100 kpc. They conducted their study exploring galaxies from z=2 to now 
using three different semi-analytical models \citep{Bower2006,DeLucia2007, Guo2011}. 
At z=0, the theoretical expectations suggested
that the fraction of massive galaxies with at least a satellite down to 1:10 ranges from 
0.3 to 0.4 and down to 1:100 from 0.6 to 0.7. We have studied that fraction in our work 
once applied the clustering corrections. E type (0.28, 0.63), 
S0 type (0.10,0.39), Sa type (0.14, 0.35), Sb/c type (0.06, 0.20) for 1:10 and 1:100, respectively.
For the full sample of massive galaxies, we get 0.14 studying satellites up to 1:10 and 
0.39 down to 1:100. Summarizing, the theory makes a good prediction if we compare their 
numbers with massive ellipticals but it obtains an important over-prediction of satellites
when we extend the host sample to all our host galaxies.

It is worth stressing that this discrepancy among the theoretical and the observational 
results can not be explained due to the different volumes of exploration used in 
both works: a spatial sphere of 100 kpc in ~\citet{Quilis2012} and a cylinder in 
redshift here. Because the way we have selected our galaxies in redshift, basically
all the satellites in the line of sight of the host within a projected 
radial distance up to 100 kpc are taken. In that sense, at comparing with the 
Millennium Simulation our number of observed satellites should be an upper limit
(as they are only restricted to 100 kpc in depth). As the number of theoretical
satellites is larger than observed, we can confidently claim  that there is a 
discrepancy with our observations. Finally, a potential loss of satellites in 
the work by ~\citet{Quilis2012} due to resolution effects will also increase
the discrepancy between the simulations and the observations.

In this same theoretical context, we find the work of \citet{QuanGuo2013}
who investigated the luminosity functions of galactic 
satellites around isolated bright hosts from model satellites placed into
the Millennium and Millennium II dark matter simulations by the GALFORM 
semi-analytic galaxy formation model.
These luminosity functions allow us to compare with our results. 
In their work, \citet{QuanGuo2013} used bright hosts within the range $-21.5<M_r<-22.5$
and the same search radius than us. They found that the number of satellites per
host was 0.29 and 0.07 (see the upper panel of their fig.8) 
segregating their sample in red and blue hosts for $\Delta m_r\sim$2.5, respectively. 
Around this value, we find that the abundance of satellites is 0.15-0.21 for (Sa, Sb/c) galaxies.
Our number of satellites per red host is also much larger than the found by \citet{QuanGuo2013}.
We obtain 0.41-0.60 for $\Delta m_r\sim$2.5. 
For less brighter satellites, $\Delta m_r\sim$4, \citet{QuanGuo2013} obtain for red and 
blue hosts 0.5 and 0.11, respectively, whereas we obtain the ranges 0.40-0.72
and 0.18-0.29 for (E, S0) and (Sa, Sb/c), respectively. The abundance of satellites for our 
blue hosts remains a factor of 2 larger whereas the number of satellites found by 
\citet{QuanGuo2013} increase significantly to reach a similar amount to ours.
In their work, \citet{QuanGuo2013} also compared with observational results from SDSS DR8.
Despite having found a good agreement between observations and models for red primaries, 
they found dramatic differences around blue primaries, placing the model a factor of 2-3
fewer satellites than are present around comparable SDSS primaries. 

\subsection{Mass and efficiency of the dark matter haloes}

Assuming that there is a link between the dark matter halo mass and the
number of satellites a galaxy has \citep{Wang2012,Wang2013,Kawinwanichakij2014}, 
we can speculate how the relative abundance of satellites found in our
different morphological samples are related to their halo masses. 
For example, if we compare the combined abundance of satellites around E
and S0 types down to 1:100, with the number of satellites per host around Sb/c massive
galaxies (grouping early versus late type), we obtain a factor of 3 higher in 
the number of satellites around early-type galaxies than late types. This
implies that on average, the dark matter halo mass of early massive galaxies
could be three times larger than the halo masses associated with our Sb/c galaxies.
We assume here that the number of satellites is approximately proportional to
the dark matter halo mass \citep{Wang2012,Wang2013}. \citet{Wang2012} also studied that 
difference using massive galaxies (log(M$_\star$/M$_{\sun}$)$\sim$11.2). In their
work, they found that the red centrals had about a factor of 2 more satellites 
than blue centrals.

Interestingly, \citet{Mandelbaum2006} estimated the efficiency with which
baryons in the halo of the galaxies have been converted into stars using a
large sample of weak gravitational lenses (0.02$<$z$<$0.3), finding that the
relative efficiency between late- and early-type galaxies with stellar masses
(see their table 3) between log(M$_\star$/M$_{\sun}$)$\sim$11.0 and 11.3 (the mean stellar
mass of our samples is log(M$_\star$/M$_{\sun}$)$\sim$11.11) was 2.5-4.36. If we consider a
similar stellar mass for our samples and using the Equation 7 of 
\citet{Mandelbaum2006}, we can compare our relative conversion efficiency.
That range of relative efficiency between late and early types is, considering our 
background and clustering corrections, 2.6 and 2.9, a value in
good agreement with their results.
Following the above discussion, we can claim that the haloes of early-type 
massive galaxies are typically 2-3 times less efficient to convert the baryons 
into stars than the haloes of late-type massive galaxies. A factor which rises, in our more 
extreme comparison, to 3-4 at comparing the sample of ellipticals and the Sb/c types. 
At extending that relative efficiency in the terms explained before, 
to other morphological types, we find that our massive late spirals (Sa) are 15-43 per
cent less efficient than the Sb/c ones. Within early-type ones, the massive
lenticulars are a 58-74 per cent more efficient producing stars than ellipticals ones.

It is also interesting to compare with the statistical approach carried out by
\citet{Moster2010} to determine the relationship between the stellar masses of
galaxies and the masses of the dark matter haloes in which they reside. In their 
work, they estimated the average number of satellites as a function of the 
dark matter halo mass using a halo occupation model. The mean number of 
satellites is illustrated in their central panel of their fig.10.
Under the assumption that down to 300 kpc we are taking all the satellites around
our massive galaxies and, considering that our completeness is around log($M$/M\sun)$\sim$9.0
( pointed line in their fig.10), we could do a direct comparison to establish the typical
halo mass associated with each one of our samples using the abundance of satellites obtained
after the corrections of background and clustering. We obtain for E-type massive 
galaxies a halo mass M$_{\rm Halo}\sim$8.5-14.9x10$^{12}$M$_{\sun}$, for the 
S0 types M$_{\rm Halo}\sim$5.4-7.2x10$^{12}$M$_{\sun}$, for the Sa types 
M$_{\rm Halo}\sim$3.2-4.5x10$^{12}$M$_{\sun}$ and for Sb/c types a
halo mass M$_{\rm Halo}\sim$2.2-3.5x10$^{12}$M$_{\sun}$.

From these data, we can conduct other interesting estimation using the 
average stellar mass of our samples of massive galaxies and the equation (7) from 
\citet{Mandelbaum2006}. This allows us to estimate again the conversion efficiency 
of baryons $\eta$ obtained for the different morphological types E, S0, Sa and
Sb/c starting from the \citet{Moster2010}'s results. We find
0.06-0.10, 0.13-0.17, 0.19-0.27 and 0.24-0.39, respectively.  Interestingly, at combining (E,S0)
and (Sa,Sb/c), grouping in early and late-types, we find a relative 
conversion efficiency of 2.3-2.4 between late and early-type galaxies. A factor 
close to the one already showed before when we compare that conversion
efficiency with the \citet{Mandelbaum2006}'s work. In this context, 
\citet{More2011} find that the difference between halo mass of red and 
blue centrals is $\sim$0.4 dex as the stellar mass of the central is 
log(M$_\star$/M$_{\sun}$)$\sim$11.1. This difference is also computed by 
\citet{Kawinwanichakij2014} and \citet{Phillips2014}. They find a lower factor
studying samples of quiescent and star-forming massive galaxies with 
log(M$_\star$/M$_{\sun}$)$>$10.78 (1$<$z$<$3) and 
log(M$_\star$/M$_{\sun}$)$\sim$ 10.5 (locally). Their samples of quiescent
centrals have a higher median halo mass by a factor of $\sim$0.3 dex (factor 2).
This lower factor compared to our results and the previous results found
in the literature could be produced by the combination of two factors, a lower mean 
stellar mass of the galaxies explored, and a larger difference among the 
halo masses of the samples studied in the nearby Universe respect to the
halo masses of the galaxies in the redshift range window of \citet{Kawinwanichakij2014}'s sample.

Finally, \citet{Dutton2010} estimated the star formation efficiencies from
satellite kinematics, weak gravitational lensing, and halo abundance matching
at redshift z$\sim$0. They found that the formation efficiency of early-type 
galaxies reached a peak of $\sim$12 per cent at M$_{\star}\sim$10$^{10.5}$h$^{-2}$M$_{\sun}$, 
decreasing to 2.8 per cent at M$_{\star}\sim$10$^{11.4}$h$^{-2}$M$_{\sun}$. In contrast, this 
efficiency was between 26 and 33 per cent for late-type galaxies whose stellar mean mass 
estimated was M$_{\star}\sim$~10$^{11.0}$h$^{-2}$M$_{\sun}$. Both results are consistent 
with those ones showed before and therefore reinforce our hypothesis of proportionality 
between number of satellites and dark mass halo we assume as a starting point.

\subsection{The main contributor to the growth of massive galaxies}

There is growing consensus ~\citep[see e.g. a discussion in][]{Trujillo2011} 
that the size evolution can not be entirely explained by internal mechanisms 
like active galactic nuclei (AGN)  
activity \citep[][]{Fan2008,Fan2010,Ragone2011}. However, it is not clear what is
the relevance of major versus minor merging in the growth of the galaxies. 
On one hand, major mergers \citep[e.g.][]{Ciotti2001,Nipoti2003,Boylan-Kolchin2006,Naab2007} 
seem to be very scarce ~\citep[at least since z$\sim$1;][]{Bundy2009,deRavel2009,
Wild2009,Lopez-Sanjuan2010,Kaviraj2011}  to play a major role in the growth of the galaxies. 
On the other hand, minor merging ~\citep[favoured theoretically for its efficiency
on increasing the size of the galaxies;][]{Khochfar-Burkert2006,
Maller2006,Hopkins2009b,Naab2009} confronts some problems with
the number of satellites found at z$\sim$1 ~\citep[e.g.][]{Ferreras2014}. 

In order to investigate the physical origin of the observed strong increase in galaxy sizes
since redshift z$\sim$2, \citet{Oser2012} led a theoretical study in which they found that
the evolution of massive early-type galaxies and their present-day properties are predominantly 
determined by frequent mergers of moderate mass (1:5) and not only by major mergers. 
\citet{Ferreras2014} probed the merging channel of massive galaxies 
(M$_\star \gtrsim$10$^{11}$M$\odot$) over the z=0.3-1.3 redshift window and down to a 
mass ratio satellite-host 1:100, segregating their sample into early and late-type
massive galaxies (see their fig.9). They found that the main contributor to 
the growth of the host mass is those satellites whose mass ratio satellite--host 
is $\sim$1:3 for both samples.

We find at z=0 a merger channel dominated by satellites with 
mass ratio satellite--host larger than 1:5 for E-type massive galaxies, or
dominated by satellites more massive than 1:2 if we consider our clustering 
correction as the most reliable. We find same result for the rest of morphological types,
the most massive satellites are the main mass growth contributors. 
The growth of massive S0, Sa and Sb/c galaxies seems to be dominated by 
satellites more massive than 1:2. However, it is evident that massive 
ellipticals have a significant large number of satellites with masses ranging 
between 1:2 and 1:5. Concretely, and if we focus on our results after clustering correction, 
the contribution of these satellites is similar to the ones
within the range 1:1-1:2. This is not seen in the rest of morphological types. The S0 types also show
a non-negligible contribution to the host mass within the interval 1:2-1:5 but lower than
a factor of 2 compared to the contribution of satellites of similar mass around ellipticals. 
In contrast, spiral types barely have satellites in this mass range. It is then interesting to study 
if this 1:2-1:5 merging channel remains growing when we assess the contribution of the 
satellites around a still more massive host sample of massive ellipticals and if it is 
reproduced at higher redshift. In this context, a better identification of massive 
ellipticals and lenticulars may be key to check whether this highlighted channel 1:2-1:5 also exists.

As the merger channel since z$\sim$1 is similar to the one found here locally, 
the observations suggest that the mass and size 
increase of the elliptical massive galaxies will be dominated by satellites with mass 
ratio within 1:1-1:5 ~\citep[see also][]{Lopez-Sanjuan2012, Ruiz2014} whereas other
morphological types as S0, Sa and Sb/c seem to associate that increase to mergers
with satellites with similar masses to the host. Note, however, that these
statements assume that the merger time-scale are independent of the mass ratio between the satellites
and the host galaxies. More realistic scenarios ~\citep[e.g.][]{Jiang2013} suggest that the merger 
time-scale rises as the mass ratio between both galaxies increases. Accordingly, the smaller 
satellites will take significantly more time to merge with the host galaxies than the more massive ones. 

Based on this, what we can claim with some confidence is that low-mass satellites with mass ratio below
1:10 would play a minor role in the mass increase of the host galaxies. They would be just very small 
in number to contribute to the mass growth, plus they will have very large time-scales to efficiently
infall into the massive galaxies. However, the small satellites could be playing a major role 
in the construction of the stellar haloes of the galaxies \citep[see e.g.][]{Cooper2013}.
If the theoretical expectation remains, and the merger time-scales 
are shorter for the most massive satellites, this mass growth due to the larger satellites will be
even more important than the result showed in Fig.~\ref{figure3}. 

\section{Summary and conclusions}\label{sec:conclusions}

In this paper we explore the abundance of satellites around 254 massive 
(10$^{11}<$M$_\star<$2$\times$10$^{11}$M$\odot$) low-$z$ (z $<$ 0.025) galaxies
visually classified as E, S0, Sa and Sb/c. 
Using the SDSS DR10 spectroscopic catalogue, the proximity of our host galaxies
guarantees that we can explore satellites with completeness down to 
M$_\star\sim$ 10$^{9}$M$\odot$. Our satellite galaxies have been 
identified within a projected radial distance of 300 kpc around the central galaxy. 
A careful statistical analysis of the background and clustering has been applied to 
decontaminate the number of satellites from fake satellites.
The abundance of satellites decline significantly from the E galaxies to S0, Sa and Sb/c types
showing an important dependence with the morphology of the host independently of 
the mass ratio satellite--host. The average number of satellites down to a mass 
ratio 1:100 within 300 kpc is $4.5\pm0.3$ for E hosts, $2.6\pm0.2$
for S0, $1.5\pm0.1$ for Sa and $1.2\pm0.2$ for Sb/c. These 
quantities decrease by a factor of 2.5-3.5 down to 1:10. 

Under the assumption that there is a proportionality between the number of satellites found and the 
dark matter halo mass, we find that the haloes of massive ellipticals are less efficient than
the haloes of Sb/c types on converting their baryons into stars by a factor of 3-4. We need
a halo 3-4 times more massive to create an elliptical with the same stellar mass than a 
Sb/c spiral. This factor decreases to $\sim$3 when we group our samples in early- and late-type massive galaxies.

If the satellites would eventually infall into their host galaxies, the growth of massive galaxies 
will be dominated by satellites with a mass ratio down to 1:10. Those satellites are the main
contributors to the stellar mass enclosed by the satellites and responsible of the 
67.2, 68.4, 88.1 and 85.7 per cent of the total mass in satellites 
for E, S0, Sa, and Sb/c types, respectively, down to 300 kpc. Massive ellipticals seem to be surrounded by a 
remarkably larger number of poor massive satellites whereas the rest of morphological types typically 
merge with more massive objects. Specifically, the main contributor to the growth of massive
spirals seems to be the satellites more massive than 1:2 (to the S0, Sa and Sb/c types). To the E hosts,
the merger channel peaks within 1:1-1:5, in agreement with the 1:5 pointed by ~\citet{Oser2012}.

These results could be used in future works to test the $\Lambda$CDM predictions about the 
number of satellites surrounding the most massive galaxies in the present-day Universe according
to their morphology. In addition, the results presented here show the most likely merging channel
of present-day massive galaxies. Finally, our work highlights
the importance of the environment where massive galaxies are immersed and how that environment is 
strongly linked to the host morphology.

\section*{Acknowledgements}

We thank the referee for detailed and constructive revision
of the manuscript. This work has been supported by the 
`Programa Nacional de Astronom\'{\i}a y Astrof\'{\i}sica' of the
Spanish Ministry of Science and Innovation under grant AYA2013-48226-C3-1-P. EM-Q acknowledges 
the support of a European Research Council Consolidator Grant (PI: McLure).

This project has made use
of data from the Sloan Digital Sky Survey (SDSS). Funding for the SDSS has been provided by the Alfred
P. Sloan Foundation, the Participating Institutions, the National Science Foundation, the US
Department of Energy, the National Aeronautics and Space Administration, the Japanese Monbukagakusho,
the Max Planck Society and the Higher Education Funding Council for England. The SDSS web site is
\url{http://www.sdss.org/}

The SDSS is managed by the Astrophysical Research Consortium for the
Participating Institutions. The Participating Institutions are the American
Museum of Natural History, Astrophysical Institute Potsdam, University of
Basel, University of Cambridge, Case Western Reserve University, University of
Chicago, Drexel University, Fermilab, the Institute for Advanced Study, the
Japan Participation Group, Johns Hopkins University, the Joint Institute for
Nuclear Astrophysics, the Kavli Institute for Particle Astrophysics and
Cosmology, the Korean Scientist Group, the Chinese Academy of Sciences
(LAMOST), Los Alamos National Laboratory, the Max-Planck-Institute for
Astronomy (MPIA), the Max-Planck-Institute for Astrophysics (MPA), New Mexico
State University, Ohio State University, University of Pittsburgh, University
of Portsmouth, Princeton University, the United States Naval Observatory, and
the University of Washington.

\bibliography{spdms}
\bibliographystyle{mn2e}

\end{document}